\documentclass[journal,letter]{IEEEtran}
\usepackage[latin1]{inputenc}
\usepackage[cmex10]{amsmath}
\usepackage{amssymb}
\usepackage{mathrsfs}

\newcommand{\sett}[1]{\mathcal{#1}}

\newcommand{\set}[1]{\mathbb{#1}}

\newtheorem{theorem}{\bf{Theorem}}
\newtheorem{definition}[theorem]{\bf{Definition}}
\newtheorem{proposition}[theorem]{\bf{Proposition}}

\newtheorem{corollary}[theorem]{\bf{Corollary}}

\begin{document}
\title{Nonextensive Generalizations of the Jensen-Shannon Divergence}

\author{Andr\'{e}~F.~T.~Martins,\ \
        Pedro~M.~Q.~Aguiar,~\IEEEmembership{Senior Member,~IEEE,}\\
        and M\'{a}rio~A.~T.~Figueiredo,~\IEEEmembership{Senior Member,~IEEE}
        \thanks{
This work was partially supported by {\it Funda\c{c}\~{a}o
para a Ci\^{e}ncia e Tecnologia} (FCT), Portuguese Ministry of Science
and Higher Education, under project PTDC/EEA-TEL/72572/2006.}
\thanks{
A. Martins is with the Department of Electrical and Computer
Engineering, {\it Instituto Superior T\'{e}cnico}, 1049-001 Lisboa, Portugal,
and with the Language Technologies Institute, Carnegie Mellon University,
Pittsburgh, PA 15213, USA. Email: afm@cs.cmu.edu}
\thanks{P. Aguiar is with the {\it Instituto de Sistemas e Rob\'{o}tica} and
the Department of Electrical and Computer
Engineering, {\it Instituto Superior T\'{e}cnico}, 1049-001 Lisboa, Portugal;
email: aguiar@isr.ist.utl.pt}
\thanks{M. Figueiredo is with the {\it Instituto de Telecomunica\c{c}\~{o}es}
and the Department of Electrical and Computer
Engineering, {\it Instituto Superior T\'{e}cnico}, 1049-001 Lisboa, Portugal;
email: mario.figueiredo@lx.it.pt}}

\maketitle

\begin{abstract}
Convexity is a key concept in information theory, namely via the
many implications of Jensen's inequality, such as the
non-negativity of the Kullback-Leibler divergence (KLD).
Jensen's inequality also underlies the concept of Jensen-Shannon
divergence (JSD), which is a symmetrized and smoothed version of
the KLD. This paper introduces new JSD-type divergences,
by extending its two building blocks: convexity and Shannon's entropy.
In particular, a new concept of $q$-convexity is introduced
and shown to satisfy a Jensen's $q$-inequality. Based on this
Jensen's $q$-inequality, the Jensen-Tsallis  $q$-difference is
built, which is  a nonextensive generalization of the JSD, based
on Tsallis entropies. Finally, the Jensen-Tsallis $q$-difference
is charaterized in terms of convexity and extrema.
\end{abstract}


\IEEEpeerreviewmaketitle

\section{Introduction}
The central role played by the
Shannon entropy in information theory has
stimulated the proposal of several generalizations and extensions
during the last decades (see, {\it e.g.}, \cite{Renyi1960,
Havrda1967, Daroczy1970, Basseville1989, Kapur1994, Arndt2004, Tsallis1988}).
One of the best known of these generalizations is the family of R\'{e}nyi
entropies, which has the Shannon entropy as a limit case
\cite{Renyi1960}, and has been used in several applications ({\it
e.g.}, \cite{Baraniuk2001,Erdogmus2004}). The R\'{e}nyi and Shannon
entropies share the well-known \emph{additivity} property, under
which the joint entropy of a pair of independent random variables is
simply the sum of the individual entropies. In other
generalizations, namely those introduced by Havrda-Charv\'{a}t
\cite{Havrda1967}, Dar\'{o}czi \cite{Daroczy1970}, and Tsallis
\cite{Tsallis1988}, the additivity property is abandoned, yielding
the so-called \emph{nonextensive} entropies. These nonextensive
entropies have raised great interest among physicists in modeling
certain physical phenomena (such as those exhibiting long-range
interactions and multifractal behavior) and as a framework for
nonextensive generalizations of the classical Boltzmann-Gibbs
statistical mechanics \cite{Abe2006, AbeOkamoto2001}. Nonextensive
entropies have also been recently used in signal/image processing
({\it e.g.}, \cite{HuTungGao2006, LiFanLi2006, Martin2004}) and many
other areas \cite{GellMannTsallis2004}.

Convexity  is a key concept in information theory, namely via the
many important corollaries of Jensen's inequality \cite{Jensen1906}, such as
the non-negativity of the relative Shannon entropy, or Kullback-Leibler
divergence (KLD) \cite{Cover1991}. The Jensen inequality is also at the basis
of the concept of Jensen-Shannon divergence (JSD), which is a
symmetrized and smoothed version of the KLD \cite{Lin1990, Lin1991}.
The JSD is widely used in areas such as statistics,  machine
learning, image and signal processing, and physics.

The goal of this paper is to introduce new extensions of JSD-type
divergences, by extending its two building blocks: convexity and the
Shannon entropy. In previous work \cite{Martins2008a}, we investigate how these
extensions may be applied in kernel-based
machine learning.
More specifically, the main
contributions of this paper are:
\begin{itemize}
\item
The concept of $q$-convexity, as a generalization of convexity, for
which we prove a {\it Jensen $q$-inequality}. The related concept of
{\it Jensen $q$-differences}, which generalize Jensen differences,
is also proposed. Based on these concepts, we introduce the {\em
Jensen-Tsallis $q$-difference}, a nonextensive generalization of the
JSD, which is also a ``mutual information'' in the sense of Furuichi
\cite{Furuichi2006}.
\item
Characterization of the Jensen-Tsallis $q$-difference, with respect
to convexity and its extrema, extending results obtained by Burbea and Rao
\cite{Burbea1982} and by Lin \cite{Lin1991} for the JSD.

\end{itemize}

The rest of the paper is organized as follows. Section
\ref{sec:tsallis} reviews the concepts of nonextensive entropies,
with emphasis on the Tsallis case.
%Section \ref{sec:denormalization}
%introduces  denormalization formulae for several entropies and
%divergences.
Section \ref{sec:jensendiffdiv} discusses Jensen differences and divergences.
The concepts of $q$-differences and $q$-convexity are
introduced in Section \ref{sec:qconvexdiff}, where they are used
to define and characterize some new divergence-type quantities.
Section \ref{sec:tsallisqdiff} defines the Jensen-Tsallis $q$-difference and derives some properties.
 Finally, Section
\ref{sec:conclusions} contains  concluding remarks and mentions
directions for future research.

\section{Nonextensive Entropies}
\subsection{Suyari's Axiomatization}
\label{sec:tsallis} Inspired by the Shannon-Khinchin axiomatic
formulation of the Shannon entropy \cite{Khinchin1957, Shannon1949},
Suyari proposed an axiomatic framework for nonextensive entropies
and a uniqueness theorem  \cite{Suyari2004}. Let
\begin{equation}
\Delta^{n-1} := \left\{(p_1,\ldots,p_n) \in \set{R}^n : p_i \ge 0,
\sum_{j=1}^n p_i = 1\right\}
\end{equation}
denote the $(n-1)$-dimensional simplex.
The Suyari axioms (see Appendix) determine the function $S_{q,\phi}:\Delta^{n-1}\rightarrow \set{R}$ given by
\begin{equation}\label{eq:nonextentropy_suyariaxioms}
S_{q,\phi}(p_1,\ldots,p_n) = \frac{k}{\phi(q)}\left(1-\sum_{i=1}^n
p_i^q\right),
\end{equation}
where $q, k \in \set{R}_+$, $S_{1,\phi} := \lim_{q\rightarrow
1}S_{q,\phi}$, and $\phi:\set{R}_+ \rightarrow \set{R}$ is a
continuous function satisfying the following three conditions:
\begin{description}
\item[\emph{(i)}] $\phi(q)$ has the same sign as $q\!-\!1$;
\item[\emph{(ii)}]
$\phi(q)$ vanishes if and only if $q=1$;
\item[\emph{(iii)}] $\phi$ is
differentiable in a neighborhood of $1$ and $\phi'(1) = 1$.
\end{description}
For any
$\phi$ satisfying these conditions, $S_{q,\phi}$ has the
\emph{pseudoadditivity} property: for any two independent random
variables $A$ and $B$,  with probability mass functions $p_A \in
\Delta^{n_A}$ and $p_B \in \Delta^{n_B}$, respectively,
\[
S_{q,\phi}(A \times B) = S_{q,\phi}(A) + S_{q,\phi}(B) -
\frac{\phi(q)}{k}\, S_{q,\phi}(A)S_{q,\phi}(B),
\]
where we denote (as usual) $S_{q,\phi}(A) := S_{q,\phi}(p_A)$.

For $q=1$, we recover the Shannon entropy,
\begin{equation}
S_{1,\phi}(p_1,\ldots,p_n) = H(p_1,\ldots,p_n) = -k\sum_{i=1}^n p_i \ln
p_i,
\end{equation}
thus pseudoadditivity turns into \emph{additivity}.

\subsection{Tsallis Entropies}
Several proposals for $\phi$ have appeared \cite{Havrda1967,
Daroczy1970, Tsallis1988}. In the rest of the paper,
we set $\phi(q) = q-1$, which yields the Tsallis entropy:
\begin{equation}\label{eq:defTsallisCountable}
S_q(p_1,\ldots,p_n) = \frac{k}{q-1}\, \left(1-\sum_{i=1}^n
p_i^q\right).
\end{equation}
To simplify, we let $k=1$ and write the Tsallis entropy as
\begin{equation}
S_q(X) := S_q(p_1,\ldots,p_n) = -\sum_{x\in X} p(x)^q \, \ln_q p(x),
\label{eq:tsallisentropy}
\end{equation}
where $\ln_q(x) := (x^{1-q}-1)/(1-q)$ is the  \emph{$q$-logarithm
function}, which satisfies $\ln_q(xy)=\ln_q(x)+x^{1-q}\ln_q(y)$ and
$\ln_q(1/x)=-x^{q-1}\ln_q(x)$.

Furuichi derived some information theoretic properties of Tsallis
entropies \cite{Furuichi2006}. Tsallis \emph{joint} and
\emph{conditional entropies} are defined, respectively, as
\begin{equation}
S_q(X,Y) := -\sum_{x,y} p(x,y)^q \ln_q p(x,y)
\end{equation}
and
\begin{eqnarray}
S_q(X|Y) &:=& -\sum_{x,y} p(x,y)^q \ln_q p(x|y) \nonumber\\
&=&  \sum_y p(y)^q S_q(X|y),
\end{eqnarray}
and the chain rule $S_q(X,Y) = S_q(X) + S_q(Y|X)$ holds.

For two probability mass functions $p_X,\, p_Y \in \Delta^n$, the
\emph{Tsallis relative entropy}, generalizing the KLD, is defined as
\begin{equation}\label{eq:tsallisrelentropy}
 D_q(p_X\|p_Y) := -\sum_x p_X(x) \ln_q
\frac{p_Y(x)}{p_X(x)}.
\end{equation}
Finally, the \emph{Tsallis mutual entropy} is defined as
\begin{equation}\label{eq:tsallismutinf}
I_q(X;Y) := S_q(X) - S_q(X|Y) = S_q(Y) - S_q(Y|X),
\end{equation}
generalizing (for $q > 1$) Shannon's mutual information
\cite{Furuichi2006}. In Section~\ref{sec:tsallisqdiff}, we
establish a relationship between Tsallis mutual entropy and a
quantity called \emph{Tsallis $q$-difference}, generalizing the one
between mutual information and the JSD \cite{Grosse2002}.

Furuichi considers an alternative generalization of Shannon's mutual
information,
\begin{equation}\label{eq:tsallismutinf_alter}
\tilde{I}_q(X;Y) := D_q( p_{X,Y} \| p_X\! \otimes p_Y ),
\end{equation}
where $p_{X,Y}$ is the true joint probability mass function of
$(X,Y)$ and $p_X\! \otimes p_Y$ denotes their joint probability if
they were independent \cite{Furuichi2006}. This alternative
definition has also been used as a ``Tsallis mutual entropy''
\cite{Lamberti2003}; notice that  $I_q(X;Y) \ne \tilde{I}_q(X;Y)$ in general, the case $q=1$ being a notable exception. In
Section~\ref{sec:tsallisqdiff}, we show that this alternative
definition also leads to a nonextensive analogue of the JSD.

\subsection{Denormalization of Tsallis Entropies}
In the sequel, we extend the domain of Tsallis entropies from $\Delta_{n-1}$ to
the set of \emph{unnormalized} measures, $\set{R}_+^n := \{(x_1,\ldots,x_n) \in \set{R}^n \,\,|\,\, \forall i \,\, x_i \ge 0\}$.
The Tsallis entropy of a measure is defined as
\begin{equation} \label{eq:defTsallisUnnorm}
S_q(x_1,\ldots,x_n) := -\sum_{i=1}^n x_i^q \ln_q x_i = \sum_{i=1}^n \varphi_q(x_i),
\end{equation}
where $\varphi_q:\set{R_{+}}\rightarrow \set{R}$ is given by
\begin{equation} \label{eq:phi_entropy_nonext}
\varphi_q(y) = -y^q \ln_q y = \left\{
\begin{array}{ll}
-y \ln y, & \text{if $q=1$},\\
(y - y^q)/(q-1), & \text{if $q\ne 1$}.
\end{array}
\right.
\end{equation}

\section{Jensen Differences and Divergences}\label{sec:jensendiffdiv}

\subsection{The Jensen Difference}

Jensen's inequality states that, if $f$ is a concave
function and $X$ is an integrable real-valued random variable,
\begin{equation}\label{eq:Jensen}
f(E[X]) - E(f(X)) \geq 0.
\end{equation}

Burbea and Rao studied the difference in the left hand side of (\ref{eq:Jensen}),
with $f:=H_\varphi$, where $H_\varphi:[a,b]^n\rightarrow \mathbb{R}$ is
a concave function, called a \emph{$\varphi$-entropy}, defined as
\begin{equation}
H_\varphi(x) := -\sum_{i=1}^n \varphi(x_i), \label{eq:phientropy}
\end{equation}
 where
$\varphi:[a,b]\rightarrow \set{R}$ is convex \cite{Burbea1982}. The
result is called the {\it Jensen difference}, as
formalized in the following definition.

\bigskip

\begin{definition}\label{def:jensendiff}
The \emph{Jensen difference} $J_{\Psi}^{\pi}:\set{R}_+^{nm} \rightarrow \set{R}$ induced by
a (concave) generalized entropy $\Psi : \set{R}_+^n \rightarrow \set{R}$ and weighted by $(\pi_1,\ldots,\pi_m) \in \Delta^{m-1}$ is
\begin{eqnarray}\label{eq:jensen_diff}
J_{\Psi}^{\pi}(x_1,\ldots,x_m) &:=&   \Psi \left(\sum_{j=1}^m \pi_j \,
x_j \right)
- \sum_{j=1}^m \pi_j \Psi(x_j) \nonumber\\
&=& \Psi \left(E[X]\right) - E[\Psi(X)],
\end{eqnarray}
where both expectations are with respect to $(\pi_1,\ldots,\pi_m)$.
\end{definition}

\bigskip

In the following subsections, we consider several instances of
Definition~\ref{def:jensendiff}, leading to several Jensen-type
divergences.

\subsection{The Jensen-Shannon Divergence}

\label{rem:expdivequalsjdiff} Let $P$ be a random probability
distribution taking values in $\{p_y\}_{y = 1,\ldots,m} \subseteq \Delta^{n-1}$ according
to a distribution $\pi=(\pi_1,\ldots,\pi_m) \in \Delta^{m-1}$. (In
classification/estimation theory parlance, $\pi$ is called the prior
distribution and $p_y := p(.|y)$ the likelihood function.) Then,
\eqref{eq:jensen_diff} becomes
\begin{equation}\label{eq:jensendiff_prob}
J_\Psi^\pi(p_1,\ldots,p_m) =  \Psi \left(E[P]\right) -
E[\Psi(P)],
\end{equation}
where the expectations are with respect to $\pi$.

Let now $\Psi = H$, the Shannon entropy.  Consider the random variables
$Y$ and $X$, taking values respectively in $\sett{Y}=\{1,\ldots,m\}$ and
$\sett{X}=\{1,\ldots,n\}$, with probability mass functions $\pi(y):=\pi_y$ and $p(x) := \sum_{y=1}^m
p(x|y)\pi(y)$. Using standard notation of information theory
\cite{Cover1991},
\begin{eqnarray}\label{eq:jensenshannondiv_mutualinf}
J^{\pi}(P) := J_H^\pi(p_1,\ldots,p_m) & = & H(X)-H(X|Y)
\\ & = & I(X;Y),\nonumber
\end{eqnarray}
where $I(X;Y)$ is the mutual information between $X$ and~$Y$. Since
$I(X;Y)$ is also equal to the KLD between the joint distribution and
the product of the marginals \cite{Cover1991}, we have
\begin{equation}\label{eq:jensenshannondiv_2expr}
J^{\pi}(P)  =  H
\left(E[P]\right) - E[H(P)]  =  E[D (P \| E[P])].
\end{equation}

The quantity $J_H^\pi(p_1,\ldots,p_m)$ is called the
\emph{Jensen-Shannon divergence} (JSD) of $p_1,\ldots,p_m$, with
weights $\pi_1,\ldots,\pi_m$  \cite{Burbea1982}, \cite{Lin1991}.
%The term ``divergence'' is justified by the fact that it is always
%nonnegative and vanishes if and only if $P$ is deterministic.
Equality \eqref{eq:jensenshannondiv_2expr} allows two
interpretations of the JSD: \emph{(i)} the Jensen difference of the Shannon
entropy of $P$; or \emph{(ii)} the expected KLD between
$P$ and the expectation of $P$.

A remarkable fact is that $J^{\pi}(P) = \min_Q E[D (P \| Q)]$,
\emph{i.e.}, $Q^* = E[P]$ is a minimizer of $E[D (P \| Q)]$ with
respect to $Q$.  It has been shown that this property together with
equality \eqref{eq:jensenshannondiv_2expr} characterize the
so-called \emph{Bregman divergences}: they hold not only for
$\Psi=H$, but for any concave $\Psi$ and the corresponding Bregman
divergence, in which case $J_{\Psi}^{\pi}$ is the \emph{Bregman
information} (see \cite{Banerjee2005} for details).

When $m=2$ and $\pi = (1/2,1/2)$,  $P$ may be seen as a
random distribution whose value on $\{p_1,p_2\}$ is chosen by
tossing a fair coin. In this case, $J^{(1/2,1/2)}(P) =
J\!S(p_1,p_2)$, where
\begin{eqnarray}\label{eq:jsdivergence2distr}
\hspace{-0.8cm}J\!S(p_1,p_2)\!\! &\!=&\! \!
H\!\left(\!\frac{p_1+p_2}{2}\!\right)
- \frac{H(p_1)+H(p_2)}{2} \nonumber\\
\!\! &\!=&\! \!  \frac{1}{2}D\!\left(\! p_1 \Big\|
\frac{p_1+p_2}{2}\!\right) + \frac{1}{2}D\!\left(\!p_2 \Big\|
\frac{p_1+p_2}{2}\!\right),
\end{eqnarray}
as introduced in \cite{Lin1991}. It has been shown  that
$\sqrt{J\!S}$ satisfies the triangle inequality (hence being a
metric) and that, moreover, it is an Hilbertian metric
\cite{Endres2003}, \cite{Topsoe2000}. %This fact suggests that
%$\sqrt{J\!S}$ may be useful in kernel-based machine learning
%techniques \cite{Cuturi2005b}.

\subsection{The Jensen-R\'{e}nyi Divergence}
\label{ex:JRD} Consider again the scenario above
(Subsection~\ref{rem:expdivequalsjdiff}), now with the R\'{e}nyi $q$-entropy
\begin{equation}
R_q(p) = \frac{1}{1-q}\, \ln\! \sum_{i=1}^n p_i^{\,q}
\end{equation}
replacing the Shannon entropy. The R\'{e}nyi $q$-entropy is concave for $q
\in [0,1)$ and has the Shannon entropy as the limit when $q \rightarrow
1$ \cite{Renyi1960}. Letting $\Psi = R_q$,  \eqref{eq:jensendiff_prob} becomes
\begin{equation}\label{eq:jensenrenyidiv}
J_{R_q}^\pi(p_1,\ldots,p_m) =  R_q \left(E[P]\right) -
E[R_q(P)].
\end{equation}
Unlike in the JSD case, there is no counterpart of  equality
\eqref{eq:jensenshannondiv_2expr} based on the R\'{e}nyi $q$-divergence
\begin{equation}
D_{R_q}(p_1 \| p_2) = \frac{1}{q-1} \ln \sum_{i=1}^n p_{1i}^q \; p_{2i}^{1-q}.
\end{equation}

The quantity $J_{R_q}^\pi$ in
\eqref{eq:jensenrenyidiv} is called the \emph{Jensen-R\'{e}nyi divergence} (JRD).
Furthermore, when $m=2$ and $\pi = (1/2,1/2)$, we write
$J_{R_q}^\pi (P) = J\!R_q(p_1,p_2)$, where
\begin{equation}\label{eq:jrdivergence2distr}
J\!R_q(p_1,p_2) = R_q\left(\frac{p_1+p_2}{2}\right) -
\frac{R_q(p_1)+R_q(p_2)}{2}.
\end{equation}
The JRD has been used in several signal/image processing
applications, such as registration, segmentation, denoising, and
classification \cite{HamzaKrim2003, He2003, Karakos2007}.

\subsection{The Jensen-Tsallis Divergence}
\label{ex:jensentsallisdiv} Burbea and Rao have defined divergences
of the form \eqref{eq:jensendiff_prob} based on the Tsallis
$q$-entropy $S_q$, defined in \eqref{eq:defTsallisUnnorm}
\cite{Burbea1982}. Like the Shannon entropy, but unlike the R\'{e}nyi
entropies, the Tsallis $q$-entropy is an
instance of a $\varphi$-entropy (see \eqref{eq:phientropy}). Letting
$\Psi = S_q$, \eqref{eq:jensendiff_prob} becomes
\begin{equation}\label{eq:jensentsallisdiv}
J_{S_q}^\pi(p_1,\ldots,p_m) =  S_q \left(E[P]\right) -
E[S_q(P)].
\end{equation}
Again, like in Subsection~\ref{ex:JRD}, if we consider the Tsallis
$q$-divergence,
\begin{equation}
D_q(p_1 \| p_2) = \frac{1}{1-q} \left( 1 - \sum_{i=1}^n {p_{1i}}^q {p_{2i}}^{1-q} \right),
\end{equation}
there is no counterpart of the equality \eqref{eq:jensenshannondiv_2expr}.

The quantity $J_{S_q}^\pi$ in
\eqref{eq:jensentsallisdiv} is called the \emph{Jensen-Tsallis
divergence} (JTD) and it has also been applied in image processing
\cite{BenHamza2006}. Unlike the JSD, the JTD lacks an interpretation
as a mutual information. In spite of this, for $q \in [1,2]$, the JTD
exhibits joint convexity  \cite{Burbea1982}. In the next section, we
propose an alternative to the JTD which, amongst other features, is
interpretable as a nonextensive mutual information (in the sense of
Furuichi \cite{Furuichi2006}) and is jointly convex, for $q \in
[0,1]$.

\section{$q$-Convexity and $q$-Differences}\label{sec:qconvexdiff}
\subsection{Introduction}
This section introduces a novel class of functions, termed
\emph{Jensen $q$-differences} (JqD), that generalizes Jensen
differences. We will later (Section~\ref{sec:tsallisqdiff}) use the
JqD to define the \emph{Jensen-Tsallis $q$-difference} (JTqD), which
we will propose as an alternative nonextensive generalization of the
JSD, instead of the JTD discussed in
Subsection~\ref{ex:jensentsallisdiv}.

We begin by recalling the concept of $q$-expectation, which is used
in nonextensive thermodynamics \cite{Tsallis1988}.

\bigskip

\begin{definition} The unnormalized \emph{$q$-expectation}
of a finite random variable $X \in \sett{X}$,  with probability mass function $P_X(x)$, is
\begin{equation} E_q[X] := \sum_{x \in \sett{X}} x\, P_X(x)^q.
\end{equation}
\end{definition}

\bigskip

Of course, $q=1$ corresponds to the standard notion of expectation.
For $q \ne 1$, the $q$-expectation does not correspond to the
intuitive meaning of average/expectation ({\it e.g.}, $E_q[1] \ne 1$
in general). Nonetheless, it has been used in the construction of
nonextensive information theoretic concepts such as the Tsallis
entropy, which can be written compactly as $S_q(X)=-E_q[\ln_q p(X)]$.

\subsection{$q$-Convexity}
We now introduce the novel concept of $q$-convexity and use it to
derive a set of results, among which we emphasize  a
\emph{$q$-Jensen inequality}.
%This not$q$-convexity will be used
%later to provide necessary and sufficient conditions for the joint
%convexity of $q$-differences.

\bigskip

\begin{definition}
Let $q \in \set{R}$ and $\sett{X}$ be a convex set. A function
$f:\sett{X}\rightarrow \set{R}$ is $q$-convex if for any $x,y \in
\sett{X}$ and $\lambda \in [0,1]$,
\begin{equation}
f(\lambda x + (1-\lambda)y) \le \lambda^q f(x) + (1-\lambda)^q f(y).
\end{equation}
\end{definition}

\bigskip

Naturally,  $f$ is $q$-concave if $-f$ is $q$-convex.  Of course,
1-convexity is the usual notion of convexity. The next proposition
states the $q$-Jensen inequality.

\bigskip
\begin{proposition}\label{prop:qjensenineq}
If $f:\sett{X}\rightarrow \set{R}$ is $q$-convex, then for any $n
\in \set{N}$, $x_1,\ldots,x_n \in \sett{X}$ and $\pi =
(\pi_1,\ldots,\pi_n) \in \Delta^{n-1}$,
\begin{equation}
f\left(\sum \pi_i x_i \right) \le \sum \pi_i^q f(x_i).
\end{equation}
\end{proposition}
\begin{proof}
Use induction, exactly as in the proof of the standard Jensen
inequality ({\it e.g.}, \cite{Cover1991}).
\end{proof}

\bigskip

\begin{proposition}\label{prop:qconveximplications}
Let $f \ge 0$ and $q \ge q' \geq 0$; then,
\begin{eqnarray}
f \,\text{is $q$-convex} \,\, &\Rightarrow& f \,\text{is $q'$-convex} \label{eq:39}\\
-f \,\text{is $q'$-convex} \,\, &\Rightarrow& -f \,\text{is
$q$-convex}.\label{eq:40}
\end{eqnarray}
\end{proposition}
\begin{proof}
Implication \eqref{eq:39} results from
\begin{eqnarray*}
f(\lambda x + (1-\lambda)y) &\le& \lambda^q f(x) + (1-\lambda)^q f(y)  \\
&\le& \lambda^{q'} f(x) + (1-\lambda)^{q'} f(y),
\end{eqnarray*}
where the first inequality states the $q$-convexity of $f$ and the
second one is valid because $f(x),f(y) \ge 0$ and $t^{q'} \ge t^q
\ge 0$, for any $t \in [0,1]$ and $q \ge q'$. The proof of
\eqref{eq:40} is analogous.
\end{proof}

\subsection{Jensen $q$-Differences}
We now generalize Jensen differences, formalized in
Definition~\ref{def:jensendiff}, by introducing the concept of
Jensen $q$-differences.

\bigskip

\begin{definition}\label{def:qdiff}
For $q \ge 0$, the \emph{Jensen $q$-difference} induced by
a (concave) generalized entropy $\Psi : \set{R}_+^n \rightarrow \set{R}$ and weighted by $(\pi_1,\ldots,\pi_m) \in \Delta^{m-1}$ is
\begin{eqnarray}\label{eq:jensen_qdiff}
T_{q,\Psi}^{\pi}(x_1,\ldots,x_m) &\triangleq&   \Psi \left(\sum_{j=1}^m \pi_j \,
x_j \right)
- \sum_{j=1}^m \pi_j^q \Psi(x_j) \nonumber\\
&=& \Psi \left(E[X]\right) - E_q[\Psi(X)],
\end{eqnarray}
where the expectation and the $q$-expectation are with respect to $(\pi_1,\ldots,\pi_m)$.
\end{definition}

Burbea and Rao established necessary and sufficient conditions for
the Jensen difference of a $\varphi$-entropy to be convex
\cite{Burbea1982}. The following proposition generalizes that
result, extending it to Jensen $q$-differences.

\bigskip

\begin{proposition}\label{prop:tsallisjointconvexity}
%Let $\sett{T}$ and $\sett{X}$ be finite sets, with $|\sett{T}|=m$
%and $|\sett{X}|=n$, and let $\pi \in M_+^1(\sett{T})$.
Let
$\varphi:[0,1]\rightarrow\mathbb{R}$ be a function of class $C^2$
and consider the ($\varphi$-entropy \cite{Burbea1982}) function
$\Psi:[0,1]^n\rightarrow \mathbb{R}$ defined by $\Psi(z) :=
-\sum_{i=1}^n \varphi(z_i)$. Then, the $q$-difference
$T_{q,\Psi}^{\pi}:[0,1]^{nm}\rightarrow \mathbb{R}$ is convex if and
only if $\varphi$ is convex and $-1/\varphi''$ is $(2-q)$-convex.
\end{proposition}
\begin{proof}
The case $q=1$ corresponds to the Jensen difference and was proved
by Burbea and Rao  (Theorem~1 in \cite{Burbea1982}). Our proof extends
that of Burbea and Rao to $q\neq 1$.

In general, $y = \{y_1,...,y_m\},$ where $y_t =
\{y_{t1},...,y_{tn}\}$, thus
\begin{eqnarray*}
T_{q,\Psi}^{\pi}(y) & = & \Psi\left(\sum_{t=1}^m \pi_t y_t\right) -
\sum_{t=1}^m \pi_t^q\; \Psi(y_t) \\
& = &  \sum_{i=1}^n \left[ \sum_{t=1}^m \pi_t^q \varphi(y_{ti}) -
\varphi\left(\sum_{t=1}^m \pi_t y_{ti}\right)\right],
\end{eqnarray*}
showing that it suffices to consider $n=1$, {\it i.e.},
\begin{equation}
T_{q,\Psi}^{\pi}(y_1,\ldots,y_m) = \sum_{t=1}^m \pi_t^q \,
\varphi(y_t) - \varphi\left(\sum_{t=1}^m \pi_t y_t\right);
\end{equation}
this function is convex on $[0,1]^m$ if and only if, for every fixed
$a_1,\ldots,a_m \in [0,1]$, and $b_1,\ldots,b_m \in \set{R}$, the
function
\begin{equation}
f(x) = T_{q,\Psi}^{\pi}(a_1 + b_1 x,\ldots,a_m + b_m x)
\end{equation}
is convex in $\{x \in \set{R} : a_t + b_t x \in [0,1],\,
t=1,\ldots,m\}$. Since $f$ is $C^2$, it is convex if and only if
$f''(t) \geq 0$.

We first show that convexity of $f$ (equivalently of
$T_{q,\Psi}^{\pi}$) implies convexity of $\varphi$. Letting $c_t =
a_t + b_t x$,
\begin{equation}\label{eq:secderivjconvex}
f''(x) = \sum_{t=1}^m \pi_t^q\, b_t^2\, \varphi''(c_t) - \left(
\sum_{t=1}^m \pi_t\, b_t\right)^2 \varphi''\left(\sum_{t=1}^m
\pi_t\, c_t\right).
\end{equation}
By choosing $x=0$, $a_t = a \in [0,1]$, for $t=1,...,m$, and
$b_1,\ldots,b_m$ satisfying $\sum_t \pi_t b_t = 0$ in
\eqref{eq:secderivjconvex}, we get
\[
f''(0) = \varphi''(a)\sum_{t=1}^m \pi_t^q b_t^2,
\]
hence, if $f$ is convex, $\varphi''(a) \ge 0$ thus $\varphi$ is
convex.

Next, we show that convexity of $f$ also implies $(2-q)$-convexity
of $-1/\varphi''$. By choosing $x=0$ (thus $c_t = a_t$) and $b_t =
\pi_t^{1-q}(\varphi''(a_t))^{-1}$, we get
\begin{eqnarray}
f''(0) & = & \sum_{t=1}^m  \frac{\pi_t^{2-q}}{\varphi''(a_t)}  -
\left( \sum_{t=1}^m \frac{\pi_t^{2-q}}{\varphi''(a_t)} \right)^2
\varphi''\left(\sum_{t=1}^m \pi_t a_t\right) \nonumber \\
&=& \left( \sum_{t=1}^m  \frac{\pi_t^{2-q}}{\varphi''(a_t)}\right)
\varphi''\left(\sum_{t=1}^m \pi_t a_t\right) \nonumber \\
&& \times \left[ \frac{1}{\varphi''\left(\sum_{t=1}^m \pi_t
a_t\right)} - \sum_{t=1}^m  \frac{\pi_t^{2-q}}{\varphi''(a_t)}
\right],\nonumber
\end{eqnarray}
where the expression inside the square brackets is the Jensen
$(2-q)$-difference of $1/\varphi''$ (see
Definition~\ref{def:qdiff}). Since $\varphi''(x) \ge 0$, the factor
outside the square brackets is non-negative, thus the Jensen
$(2-q)$-difference of $1/\varphi''$ is also nonnegative and
$-1/\varphi''$ is $(2-q)$-convex.

Finally, we show that if $\varphi$ is convex and $-1/\varphi''$ is
$(2-q)$-convex, then  $f''\ge 0$, thus $T_{q,\Psi}^{\pi}$ is
convex.
%We want to show \eqref{eq:secderivjconvex} for arbitrary
%fixed $x_i \in [0,1]$, $a_i \in \set{R}$ with $x_i+a_i t \in [0,1]$,
%$i=1,\ldots,k$.
Let $r_t = (q\pi_t^{2-q}/\varphi''(c_t))^{1/2}$ and $s_t = b_t
(\pi_t^{q} \varphi''(c_t)/q)^{1/2}$; then, non-negativity of $f''$
results from the following chain of inequalities/equalities:
\begin{eqnarray}\hspace{-0.7cm}
0\hspace{-0.2cm} &\le&\hspace{-0.2cm} \left(\sum_{t=1}^m
r_t^2\right)\left(\sum_{t=1}^m s_t^2\right) -
\left(\sum_{t=1}^m r_t \, s_t\right)^2\label{eq:45}\\
\hspace{-0.2cm} &=&\hspace{-0.2cm} \sum_{t=1}^m
\frac{\pi_t^{2-q}}{\varphi''(c_t)}
\sum_{t=1}^m b_t^2 \pi_i^{q} \varphi''(c_t) -
\left(\sum_{t=1}^m b_t \pi_t\right)^2  \label{eq:quarenta_e_seis}\\
\hspace{-0.2cm} &\le&\hspace{-0.2cm} \frac{1}{\varphi''\left(
\sum_{t=1}^m \pi_t c_t \right)}
 \sum_{t=1}^m b_t^2 \pi_t^{q} \varphi''(c_t) -
 \left(\sum_{t=1}^m b_t \pi_t\right)^2  \label{eq:quarenta_e_sete}\\
\hspace{-0.2cm} &=&\hspace{-0.2cm} \frac{1}{\varphi''\left(
\sum_{t=1}^m \pi_t c_t \right)} \cdot
f''(t),\label{eq:quarenta_e_oito}
\end{eqnarray}
where: \eqref{eq:45} is the Cauchy-Schwarz inequality; equality
\eqref{eq:quarenta_e_seis} results from the definitions of $r_t$ and
$s_t$ and from the fact that $r_t s_t = b_t \pi_t$; inequality
\eqref{eq:quarenta_e_sete} states the $(2-q)$-convexity of
$-1/\varphi''$; equality \eqref{eq:quarenta_e_oito} results from
\eqref{eq:secderivjconvex}.
\end{proof}

\section{The Jensen-Tsallis $q$-Difference} \label{sec:tsallisqdiff}
\subsection{Definition}
As in Subsection~\ref{rem:expdivequalsjdiff}, let $P$ be a random
probability distribution taking values in $\{p_y\}_{y = 1,\ldots,m}$ according to
a distribution $\pi=(\pi_1,\ldots,\pi_m) \in \Delta^{m-1}$. Then, we may write
\begin{equation}\label{eq:tsallisqdiff_prob}
T_{q,\Psi}^\pi(p_1,\ldots,p_m) =  \Psi \left(E[P]\right) -
E_q[\Psi(P)],
\end{equation}
where the expectations are with respect to $\pi$. Hence Jensen
$q$-differences may be seen as deformations of the standard Jensen
differences \eqref{eq:jensendiff_prob}, in which the second
expectation is replaced by a $q$-expectation.

Let now $\Psi = S_q$, the nonextensive Tsallis $q$-entropy.
Introducing the random variables $Y$ and $X$, with values
respectively in $\sett{Y}=\{1,\ldots,m\}$ and $\sett{X}=\{1,\ldots,n\}$, with probability mass functions $\pi(y) := \pi_y$
and $p(x) := \sum_{y=1}^m p(x|y)\pi(y)$, we have (writing
$T^\pi_{q, S_q}$ simply as $T^\pi_{q}$)
\begin{eqnarray}
T_q^\pi(p_1,\ldots,p_m) = S_q(X)-S_q(X|Y) =
I_q(X;Y),\label{eq:Tq}
\end{eqnarray}
where $S_q(X|Y)$ is the Tsallis  conditional $q$-entropy, and
$I_q(X;Y)$ is the Tsallis mutual $q$-entropy, as defined by Furuichi
\cite{Furuichi2006}. Observe that \eqref{eq:Tq} is a nonextensive
analogue of \eqref{eq:jensenshannondiv_mutualinf}. Since,  in
general, $I_q\ne \tilde{I}_q$ (see \eqref{eq:tsallismutinf_alter}),
unless $q=1$ ($I_1=\tilde{I}_1=I$), there is no counterpart of
\eqref{eq:jensenshannondiv_2expr} in terms of $q$-differences.
Nevertheless, Lamberti and Majtey have proposed a non-logarithmic
version of the JSD, which corresponds  to using $\tilde{I}_q$ for
the Tsallis mutual $q$-entropy (although this interpretation is  not
explicitally mentioned by those authors) \cite{Lamberti2003}.

We call the
quantity $T_q^\pi(p_1,\ldots,p_m)$ the
\emph{Jensen-Tsallis $q$-difference} (JTqD) of $p_1,\ldots,p_m$ with
weights $\pi_1,\ldots,\pi_m$. Although the JTqD
is a generalization of the \emph{Jensen-Shannon
divergence},  for $q \ne 1$, the term ``divergence'' would be
misleading in this case, since $T_q^\pi$ may take negative values
(if $q < 1$) and does not vanish in general if $P$ is deterministic.

When $m=2$ and $\pi = (1/2,1/2)$, define $T_q :=
T_q^{1/2,1/2}$,
\begin{equation}\label{eq:tsallisdiff2distr}
T_q(p_1,p_2) = S_q\left(\frac{p_1+p_2}{2}\right) - \frac{S_q(p_1)+S_q(p_2)}{2^q}.
\end{equation}
Notable cases arise for particular values  of $q$:
\begin{itemize}
\item For $q=0$,  $S_0(p) = -1+\|x\|_0$, where $\|x\|_0$ denotes the
so-called {\it 0-norm} (although it's not a norm) of vector $x$,
{\it i.e.}, its number of nonzero components. The Jensen-Tsallis
$0$-difference is thus
\begin{equation}\label{eq:cardinaldiff}
T_0(p_1,p_2) = 1 - \|p_1 \odot p_2\|_0,
\end{equation}
where $\odot$ denotes the Hadamard-Schur ({\it i.e.}, elementwise)
product. We call $T_0$ the \emph{Boolean difference}.
\item For $q=1$, since $S_1(p) = H(p)$,  $T_1$ is the JSD,
\begin{equation}\label{eq:jsdiv1diff}
T_1(p_1,p_2) = J\!S(p_1, p_2).
\end{equation}
\item For $q=2$,  $S_2(p)=1-\langle p,p\rangle$, where $\langle x,
y \rangle = \sum_i x_i\, y_i$ is the usual inner product between  $x$ and
$y$. Consequently, the Tsallis $2$-difference is
\begin{equation}\label{eq:lineardiff}
T_2(p_1,p_2) = \frac{1}{2} - \frac{1}{2}\; \langle p_1, p_2 \rangle,
\end{equation}
which we call the \emph{linear difference}.
\end{itemize}

\subsection{Properties of the JTqD}
This subsection presents results regarding convexity and extrema of
the JTqD, for several values of $q$,  extending known properties of
the JSD ($q=1$).

Some properties of the JSD are lost in the transition to
nonextensivity. For example, while the former is nonnegative and
vanishes if and only if all the distributions are identical, this is
not true in general with the JTqD. Nonnegativity of the JTqD is only
guaranteed if $q \ge 1$, which explains why some authors ({\it
e.g.}, \cite{Furuichi2006}) only consider values of $q\ge 1$, when
looking for nonextensive analogues of Shannon's information theory.
Moreover, unless $q=1$, it is not generally true that
$T_q^{\pi}(p,\ldots,p) = 0$ or even that $T_q^{\pi}(p,\ldots,p, p')
\ge T_q^{\pi}(p,\ldots,p, p)$. For example, the solution to the optimization problem
\begin{equation}
\min_{p_1\in\Delta^n} T_q(p_1,p_2),\label{eq:minp1p2}
\end{equation}
is, in general, different from $p_2$, unless if $q=1$. Instead, this
minimizer is closer to the uniform distribution, if $q \in [0,1)$,
and closer to a degenerate distribution, for $q \in (1,2]$. This is
not so surprising: recall that $T_2(p_1,p_2) =
\frac{1}{2}-\frac{1}{2}\langle p_1,p_2 \rangle$; in this case,
\eqref{eq:minp1p2} becomes a linear program, and the solution is not
$p_2$, but $p_1^* = \delta_j$, where $j =
\arg\max_i p_{2i}$.

We start by recalling a basic result, which essentially confirms that
Tsallis entropies satify one of the Suyari axioms (see Axiom A2 in the Appendix), which states that entropies should be maximized by uniform distributions.

\bigskip
\begin{proposition}\label{prop:uniformTsallis}
%Let $\sett{X}$ be a finite set.
The uniform distribution  maximizes
the Tsallis entropy for any $q \ge 0$.
\end{proposition}
\begin{proof}
Consider the problem
\[
\max_p S_q(p), \;\;\;\mbox{subject to $\sum_i p_i = 1$ and $p_i \ge
0$.}
\]
Equating the gradient of the Lagrangian to zero, yields
\[
\mbox{$\frac{\partial}{\partial p_i} \left( S_q(p) + \lambda (\sum_i
p_i - 1)\right) = -q(q-1)^{-1}p_i^{q-1} + \lambda = 0$,}
\]
for all $i$. Since all these equations are identical, the solution
is the uniform distribution, which is a maximum, due to the
concavity of $S_q$.
\end{proof}

\bigskip

The following corollary of
Proposition~\ref{prop:tsallisjointconvexity} establishes the joint
convexity of the JTqD, for $q \in [0,1]$. This complements the joint
convexity of the JTD, for $q \in [1,2]$, which was proved by Burbea
and Rao \cite{Burbea1982}.

\bigskip
\begin{corollary}\label{cor:convextsallis}
%Let $\sett{T}$ and $\sett{X}$ be finite sets with cardinalities $m$
%and $n$, respectively.
For $q \in [0,1]$, the JTqD  is a jointly
convex function on $\Delta^{n-1}$.
%$M_+^{1,S_q}(\sett{X})$.
Formally, let $\{ p_y^{(i)}
\}_{y = 1,\ldots, m}^{i=1,\ldots,l}$, be a collection of $l$
sets of probability distributions on $\mathcal{X}=\{1,\ldots,n\}$; then, for any
$(\lambda_1,\ldots,\lambda_l) \in \Delta^{l-1}$,
\[
T_q^\pi \left( \sum_{i=1}^l \lambda_i p_{1}^{(i)}, \ldots,
\sum_{i=1}^l \lambda_i p_{m}^{(i)} \right) \le \sum_{i=1}^l
\lambda_i T_q^\pi (p_{1}^{(i)},\ldots, p_{m}^{(i)}).
\]
\end{corollary}
\begin{proof}
Observe that the Tsallis entropy \eqref{eq:tsallisentropy} of a
probability distribution $p_t = \{p_{t1},...,p_{tn}\}$ can be
written as
\[
S_q(p_t) =  - \sum_{i=1}^n \varphi(p_{ti}), \;\;\;\mbox{where}\;\;\;
\varphi_q(x) = \frac{x - x^q}{1-q};
\]
thus, from Proposition~\ref{prop:tsallisjointconvexity}, $T_q^\pi$
is convex if and only if $\varphi_q$ is convex and $-1/\varphi_q''$
is $(2-q)$-convex. Since $\varphi_q''(x) = q\,x^{q-2}$,\linebreak
$\varphi_q$ is convex for $x \geq 0$ and $q \geq 0$. To show the
$(2-q)$-convexity of $-1/\varphi_q''(x) = -(1/q) x^{2-q}$, for $x_t
\geq 0$, and $q \in [0,1]$, we use a version of the power mean
inequality \cite{Steele},
\[
-\left( \sum_{i=1}^l \lambda_i \, x_i \right)^{2-q} \le -
\sum_{i=1}^l \left(\lambda_i\, x_i \right)^{2-q} = - \sum_{i=1}^l
\lambda_i^{2-q}\, x_i^{2-q} ,
\]
thus concluding that $-1/\varphi_q''$ is in fact $(2-q)$-convex.
\end{proof}

\bigskip

The next corollary, which results from the previous one, provides an
upper bound for the JTqD, for $q\in[0,1]$. Although this result is
weaker than that of Proposition~\ref{prop:boundstsallisdiff} below,
we include it since it provides insight about the upper extrema of
the JTqD.

\bigskip
\begin{corollary}\label{cor:simplia}
Let $q \in [0,1]$. Then, $T_q^\pi(p_1,\ldots,p_m)
\le S_q(\pi)$.
\end{corollary}
\begin{proof}
From Corollary~\ref{cor:convextsallis}, for $q \in [0,1]$,
$T_q^\pi(p_1,\ldots,p_m)$ is convex. Since its domain is a convex
polytope (the cartesian product of $m$ simpleces), its maximum
occurs on a vertex, {\it i.e.}, when each argument $p_t$ is a
degenerate distribution at $x_t$, denoted $\delta_{x_t}$. In
particular, if $n
\ge m$, this maximum occurs at the vertex corresponding to
disjoint degenerate distributions, {\it i.e.}, such that $x_i \neq
x_j$ if $i\neq j$. At this maximum,
\begin{eqnarray}
T_q^{\pi}(\delta_{x_1},\ldots,\delta_{x_m}) & = &
S_q\left(\sum_{t=1}^m
\pi_t\delta_{x_t}\right) - \sum_{t=1}^m \pi_t S_q(\delta_{x_t})\nonumber\\
& = & S_q\left(\sum_{t=1}^m \pi_t\delta_{x_t}\right)\label{eq:55}\\
& = & S_q(\pi),\label{eq:56}
\end{eqnarray}
where the equality in \eqref{eq:55} results from $S_q(\delta_{x_t})
= 0$. Notice that this maximum may not be achieved if $n
< m$.
\end{proof}

\bigskip

The next proposition establishes (upper and lower) bounds for the
JTqD, extending Corollary~\ref{cor:simplia} to any non-negative $q$.
%The result still holds when $\mathcal{X}$ and $\mathcal{Y}$ are countable sets.

\bigskip
\begin{proposition}\label{prop:boundstsallisdiff}
For $q \ge 0$,
\begin{equation}
T_q^\pi(p_1,\ldots,p_m) \le S_q(\pi),\label{eq:JTqDupper}
\end{equation}
and, if $n \ge m$, the maximum is reached for a set of disjoint degenerate
distributions. As in Corollary~\ref{cor:simplia}, this maximum may
not be attained if $n < m$.

For $q\geq 1$,
\begin{equation}\label{eq:JTqDlower1}
T_q^\pi(p_1,\ldots,p_m) \geq 0,
\end{equation}
and the minimum is attained in the pure deterministic case, {\it
i.e.}, when all distributions are equal to same degenerate
distribution. Results \eqref{eq:JTqDupper} and
\eqref{eq:JTqDlower1} still hold when $\mathcal{X}$ and $\mathcal{Y}$ are countable sets.

For $q \in [0,1]$,
\begin{equation}
T_q^\pi(p_1,\ldots,p_m) \ge S_q(\pi) [1 -
n^{1-q}].\label{eq:JTqDlower}
\end{equation}
This lower bound (which is zero or negative) is attained when all
distributions are uniform.
\end{proposition}
\begin{proof}
The proof of \eqref{eq:JTqDupper}, for $q\geq 0$, results from
\begin{eqnarray}
T_q^\pi(p_1,\ldots,p_m) &=& \frac{1}{q-1}\left[ 1-\sum_j
\left( \sum_i \pi_i p_i^j \right)^q \right.\nonumber \\
&& \left. - \sum_i \pi_i^q \left( 1 - \sum_j (p_i^j)^q\right)  \right] \nonumber \\
&=& S_q(\pi) + \frac{1}{q-1} \sum_j \left[ \sum_i
\left(\pi_i p_i^j\right)^q \right.\nonumber \\
&& \left. - \left(\sum_i \pi_i p_i^j\right)^q \right]  \nonumber \\
&\le& S_q(\pi),
\end{eqnarray}
where the inequality holds since, for $y_i \geq 0$: if $q \geq 1$,
then $\sum_i y_i^q \le \left(\sum_i y_i\right)^q$; if $q \in [0,1]$,
then $\sum_i y_i^q \ge \left(\sum_i y_i\right)^q$.

The proof that $T_q^\pi \ge 0$ for $q\ge 1$, uses the  notion of
$q$-convexity. For countable $\sett{X}$, the Tsallis entropy \eqref{eq:defTsallisCountable}
is nonnegative. Since $-S_q$
is $1$-convex, then, by Proposition~\ref{prop:qconveximplications},
it is also $q$-convex for $q \ge 1$. Consequently, from the $q$-Jensen
inequality (Proposition~\ref{prop:qjensenineq}), for finite
$\sett{Y}$, with $|\sett{Y}|=m$,
\[
T_q^\pi\left(p_1,\ldots,p_m\right) = S_q\left(\sum_{t=1}^m \pi_t p_t
\right) - \sum_{t=1}^m \pi_t^q S_q(p_t)  \ge 0.
\]
Since $S_q$ is  continuous, so is $T_q^\pi$, thus the inequality is
valid in the limit as $m\rightarrow\infty$, which proves the
assertion for $\sett{Y}$ countable. Finally,
$T_q^\pi(\delta_1,\ldots,\delta_1) = 0$, where $\delta_1$ is
some degenerate distribution.

Finally, to prove \eqref{eq:JTqDlower}, for $q \in [0,1]$ and
$\sett{X}$ finite,
\begin{eqnarray}
T_q^\pi (p_1,\ldots,p_m) &=& S_q\left(\sum_{t=1}^m \pi_t p_t\right)
- \sum_{t=1}^m \pi_t^q S_q (p_t)  \nonumber \\
&\ge& \sum_{t=1}^m \pi_t S_q(p_t) - \sum_{t=1}^m \pi_t^q S_q (p_t) \label{eq:firstinequality} \\
&=& \sum_{t=1}^m (\pi_t - \pi_t^q) S_q(p_t)  \nonumber \\
&\ge& S_q(U) \sum_{t=1}^m (\pi_t - \pi_t^q)  \label{eq:secondinequality} \\
&=&  S_q(\pi) [1 - n^{1-q}],
\end{eqnarray}
where the inequality \eqref{eq:firstinequality} results from  $S_q$
being concave, and the inequality \eqref{eq:secondinequality} holds
since $\pi_t - \pi_t^q \leq 0$, for $q \in [0,1]$, and the uniform
distribution $U$ maximizes $S_q$
(Proposition~\ref{prop:uniformTsallis}), with $S_q(U) =
(1-n^{1-q})/(q-1)$.
\end{proof}

\bigskip

Finally, the next proposition characterizes the convexity/concavity
of the JTqD. As before, it holds more generally when $\sett{Y}$ and $\sett{X}$ are countable sets.

\bigskip
\begin{proposition}\label{prop:tsallisdiffconvex}
The  JTqD is convex
in each argument, for $q \in [0,2]$, and concave in each argument,
for $q \ge 2$.
\end{proposition}
\begin{proof}
Notice that the JTqD can be written as $T_q^\pi (p_1,\ldots,p_m) =
\sum_j \psi(p_{1j},\ldots,p_{mj})$, with
\begin{eqnarray}
\lefteqn{\psi(y_1,\ldots,y_m) = }\label{eq:tsallisdivsummand}\\
 && \hspace{-0.5cm} \frac{1}{q-1} \left[ \sum_i
(\pi_i - \pi_i^q) y_i + \sum_i \pi_i^q y_i^q  - \left( \sum_i \pi_i
y_i \right)^{\!q} \;\right].\nonumber
\end{eqnarray}
It suffices to consider the second derivative of $\psi$  with
respect to $y_1$. Introducing $z = \sum_{i=2}^m \pi_i\, y_i$,
\begin{eqnarray}
\frac{\partial^2 \psi}{\partial y_1^2}
&=& q \left[ \pi_1^q\, y_1^{q-2} - \pi_1^2 \, (\pi_1 \, y_1 + z )^{q-2} \right]  \nonumber \\
&=& q \, \pi_1^2 \left[ (\pi_1\, y_1)^{q-2} -
 (\pi_1\, y_1 + z)^{q-2} \;\right].\label{eq:sixtysix}
\end{eqnarray}
Since $\pi_1 \, y_1 \le (\pi_1\, y_1 + z) \le 1$,  the quantity in
\eqref{eq:sixtysix} is nonnegative for $q \in [0,2]$ and
non-positive for $q \geq 2$.
\end{proof}

\section{Conclusion}
\label{sec:conclusions}

In this paper we have  introduced new Jensen-Shannon-type divergences,
by extending its two building blocks: convexity and entropy.
We have introduced the concept of $q$-convexity, for
which we have stated and proved a {\it Jensen $q$-inequality}.
Based on this concept, we have introduced the {\em Jensen-Tsallis $q$-difference},
a nonextensive generalization of the Jensen-Shannon divergence.
We have characterized the Jensen-Tsallis $q$-difference with respect to
convexity and extrema, extending previous results obtained in
\cite{Burbea1982, Lin1991} for the Jensen-Shannon divergence.

\section*{Acknowledgments}
The authors would like to thank Prof. Noah Smith for valuable comments
and discussions.

\appendix\label{sec:suyari}

In \cite{Suyari2004}, Suyari proposed the following set of axioms (above referred as Suyari's axioms) that determine nonextensive entropies $S_{q,\phi}:\Delta^{n-1}\rightarrow \set{R}$ of the form stated in \eqref{eq:nonextentropy_suyariaxioms}. In what follows, $q$ is fixed and $f_q$ is a function defined on $\Delta^{n-1}$.
\begin{enumerate}
	\item[(A1)] \emph{Continuity}: $f_q$ is continuous in $\Delta^{n-1}$ and $q \ge 0$;
	\item[(A2)] \emph{Maximality}: For any $q\ge 0$, $n \in \set{N}$, and $(p_1,\ldots,p_n) \in \Delta^{n-1}$, $f_q(p_1,\ldots,p_n) \le f_q(1/n,\ldots,1/n)$;
	\item[(A3)] \emph{Generalized additivity}: For $i = 1,\ldots,n$, $j = 1,\ldots,m_i$, $p_{ij} \ge 0$, and $p_i = \sum_{j=1}^{m_i} p_{ij}$,
\begin{eqnarray*}
	f_q(p_{11},\ldots,p_{nm_i}) &=& f_q(p_1,\ldots,p_n) + \nonumber\\ && \sum_{i=1}^n p_i^q f_q\left( \frac{p_{i1}}{p_i}, \ldots,\frac{p_{im_i}}{p_i} \right);
\end{eqnarray*}
	\item[(A4)] \emph{Expandability}: $f_q(p_1,\ldots,p_n,0) = f_q(p_1,\ldots,p_n)$.
\end{enumerate}

\end{document}